# Physical parameter and loss determination of piezoceramics using partial electrode: $k_{31}$ and $k_{33}$ mode cases

Yoonsang Park[1,*], Hossein Daneshpajooh[1], Timo Scholehwar[2], Eberhard Hennig[2], and Kenji Uchino[1]

1) International Center for Actuators and Transducers (ICAT), The Pennsylvania State University, University Park, PA, 16802, USA

2) R&D Department, PI Ceramic GmbH, Lindenstrasse, 07589 Lederhose, Germany

**Abstract:** The standard method to determine physical parameters of piezoceramics, established by IEEE, has been utilized for decades by the number of researchers, yet it omits presence of important loss factors and possesses serious deficits that restricts accurate parameter determination. In order to resolve these issues, the partial electrode (PE) method (mechanical excitation method) was previously proposed. In this study, we aim to propose a modified PE method to enhance the efficiency of parameter determination process, along with a simplified analytical admittance equation for better understanding of the PE configuration. To prove that the PE method is reliable, possible causes of errors were listed, and it was shown that they were either negligibly small or resolved with proper calibration methods. Throughout the paper, it was validated that the PE method not only reduces the errors of several physical parameters by avoiding error propagation, but also enables measurement compatibility with commercially available impedance analyzers.

1. Introduction

High-power piezoelectric devices, such as ultrasonic transducers, actuators and voltage transformers, have been rigorously developed through recent decades[1-4]. The demand for miniaturization of high-power piezoelectric devices has grown rapidly[5,6], as the significance of micro-scale electronic devices are being emphasized. As such, piezoelectric devices have effectively replaced conventional electromagnetic counterparts with better performance and no electromagnetic noise[7-9]. As the demanding scales for electronic devices are getting smaller and smaller, further miniaturization of piezoelectric device is required. However, miniaturization of devices while maintaining energy density is still limited, due to heat generation that degrades the overall performance of high-power piezoelectric devices[7,8,10-13].

It is known that the heat generation is mainly due to "losses" in piezoelectric materials[7]. There are in general three types of losses: elastic, dielectric, and piezoelectric losses[14-16]. Those three categories are further classified as either intensive or extensive, based on specific electrical or mechanical boundary conditions. In terms of simplified 1D mathematical expression, we have[8,17,18]:

$$\varepsilon^{X*} = \varepsilon^X(1 - j\tan\delta') \tag{1}$$

$$s^{E*} = s^E(1 - j\tan\phi') \tag{2}$$

$$d^* = d(1 - j\tan\theta') \tag{3}$$

$$\kappa^{x*} = \kappa^x(1 + j\tan\delta) \tag{4}$$

$$c^{D*} = c^D(1 + j\tan\phi) \tag{5}$$

$$h^* = h(1 + j\tan\theta) \tag{6}$$

where $\varepsilon^X$ is stress (X)-constant permittivity, $s^E$ is "electric field (E)-constant" elastic compliance, $d$ is piezoelectric coefficient, $\kappa^x$ is strain (x)-constant inverse permittivity, $c^D$ is dielectric displacement ($D$)-constant elastic stiffness, and $h$ is inverse piezoelectric coefficient. The former three parameters are named "intensive physical parameters", while the latter three are called "extensive physical parameters". The primed loss tangent values in imaginary part of each complex, (with superscripted stars) are "intensive loss", whereas non-primed loss tangent values are "extensive" losses. The negative signs for intensive losses and positive signs for extensive losses are due to convention, considering the direction of loss hysteresis loop[8].

Each type of loss, either intensive or extensive, has its own significance. For example, intensive losses, as input parameters, greatly increase the accuracy of finite element analysis (FEA) computer simulation[19-21], which is a powerful tool to investigate desired targeting resonance frequency or mechanical quality factors of piezoelectric devices. Such a tool provides convenient way to design piezoelectric devices without actual fabrication. Meanwhile, extensive losses, losses in constrained boundary conditions, are helpful in elucidating heat generation mechanism due to domain wall dynamics[10,22]. Therefore, obtaining accurate values of both intensive and extensive losses are important, from both technological and scientific interests.

The standard method to determine physical parameters and losses was first established by IRE Standard[23], then further shaped by Institute for Electrical and Electronics Engineer (IEEE) in 1980s[24]. However, there are several deficits in this method that prevent users from obtaining accurate parameters. For example, according to the piezoelectric equivalent circuit (EC) described in IEEE Standard on Piezoelectricity, only one type of loss (elastic loss) is explained as a resistor in LCR circuit[24]. Therefore, until the first experimental demonstration of "piezoelectric loss" in 2000s[14], many researchers have believed that the quality factor at resonance frequency ($Q_A$) is equivalent to that at antiresonance frequency ($Q_B$). There are even more issues with IEEE Standard on piezoelectricity: for example, in $k_{33}$ mode

(a bar with electric field parallel to sound velocity) specimen, shown in Figure 1 (b), high impedance values near antiresonance frequency is the most significant issue, and there are even more issues, such as wire attachment issues and electric flux leakage[18,25-27]. Furthermore, while $k_{31}$ mode (a bar with electric field perpendicular to sound velocity) specimen, shown in Figure 1 (a), only provides intensive elastic loss originated from $Q_A$, $k_{33}$ mode, shown in Figure 1 (b), only provides extensive-like (see subsection 3.2 for further explanation) elastic loss originated from $Q_B$. For each mode, in order to obtain other types of loss values (extensive in $k_{31}$ and intensive in $k_{33}$ mode), one has to utilize either [$K$] matrix[8,10,17] or other complicated equations originated from piezoelectric constitutive relations[28], which dramatically increase error from standard deviation due to error propagation.

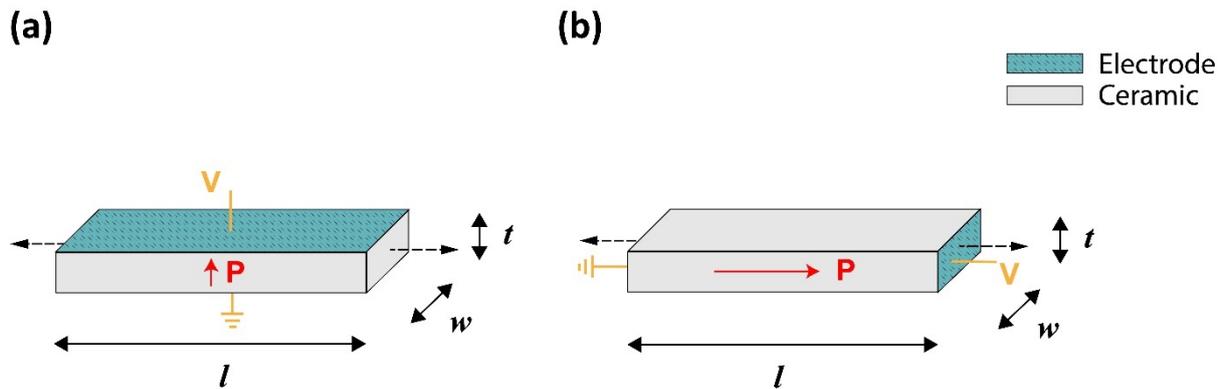

**Figure 1.** Standard (a) $k_{31}$ mode and (b) $k_{33}$ mode piezoelectric specimen. Voltage and polarization directions, as well as sample dimensions are defined. Redrawn from IEEE Standard on Piezoelectricity [20].

In order to resolve such issues, partial electrode (PE) method, which is basically a mechanical excitation method, was previously proposed[18,29], as shown in Figure 2. The PE configuration is composed of center part, which is electrically excited, and side part, which is mechanically excited by the center part. The advantage of PE is that intensive and extensive-like elastic compliances and losses can both be determined using the same configuration just by changing the surface electrode. For example, the side with no electrode, as shown in Figure 2 (c), provides extensive-like elastic compliance and loss, whereas side with electrode, as shown in Figure 2 (d), provides intensive elastic compliance and intensive elastic loss. Furthermore, since the mechanical excitation of the side specimen can be monitored by impedance/admittance measurement with the center part, experimental impedance values are in the range of $10^4$ and $10^5$ $\Omega$ for soft PZT[18,29]. Therefore, high impedance issue of the IEEE Standard $k_{33}$ mode specimen can be resolved.

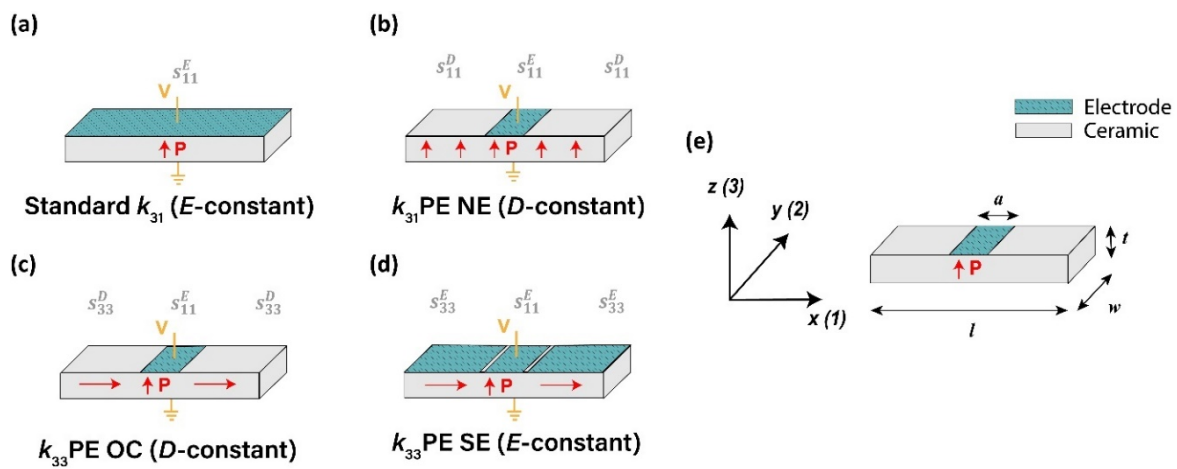

**Figure 2.** Sample geometries of (a) Standard $k_{31}$ specimen, (b) $k_{31}$ mode PE non-electrode (NE), (c) $k_{33}$ mode PE open circuit (OC) and (d) $k_{33}$ mode PE side electrode (SE) used in this study. Grey characters above each geometry denote corresponding elastic compliance to specific parts of geometry. (e) Coordinate and dimension definition of a PE sample.

In this study, we aim to provide detailed physical parameters and loss determination process using the PE method. Different from previous descriptions that included open circuit (OC) for antiresonance characterization and short circuit (SC) for resonance characterization[17,18], the number of types of PE configuration has been reduced for experimental simplicity. For determination of both intensive and extensive-like, real and imaginary elastic parameters and other physical parameters (such as dielectric and piezoelectric parameters), the following 4 types of samples are needed: IEEE Standard $k_{31}$ specimen, $k_{31}$ PE non-electrode (NE), $k_{33}$ PE open circuit (OC) and $k_{33}$ PE side electrode (SE). Throughout the paper, the following materials will be discussed: simplified and universal admittance equation for PE configuration, possible error causes for PE method and comprehensive parameter determination process using PE samples.

## 2. Material and Methods

For 4 types of samples made from both PIC 255 (Nb-doped soft PZT) and PIC 181 (Mn-doped hard PZT) [PI Ceramic GmbH, Lederhose, Germany] (soft and hard PZT for checking the samples' performance difference), 6 samples with the dimension of length ($l$) × width ($w$) × thickness ($t$) = 20 × 2.5 × 0.5 mm (See Figure 1 and 2 for dimension definition of samples) were prepared: IEEE Standard $k_{31}$ specimen, $k_{31}$ PE NE, $k_{33}$ PE OC, and $k_{33}$ PE SE. For all the samples, pure Ag was sputtered and used as electrode. For the PE samples, center electrode was maintained to about 10 % of the total $l$ of the sample, and the portion of the center part for each PE sample was measured with optical microscope. The off-resonance (for permittivity measurement) and the fundamental mode on-resonance admittance/impedance spectra for each sample were measured with 4294A Precision Impedance Analyzer [Agilent Technologies, Santa Clara, CA], with 100 mV input voltage (low vibration velocity range for escaping from the heat generation). For Standard $k_{31}$ mode, the parameter determination procedure described by Zhuang *et al.*[30] was utilized; for PE samples, experimental admittance curves were fitted to analytical equations derived in our previous work[18] to obtain elastic compliance and loss values. For each determined parameter, error was determined by using standard deviation divided by square root of measurement number, and error propagation method was utilized for parameters that were calculated through the equations. Figure 3 shows actual piezoceramic samples used in this study.

ATILA++ Finite Element Method software (distributed by Micromechatronics Inc., State College, PA) was utilized in this study, in order to verify analytical admittance equations and to observe the effect of volumes with canted polarization (see section 4.1.4). Refer to supplementary materials for more information on FEA simulation.

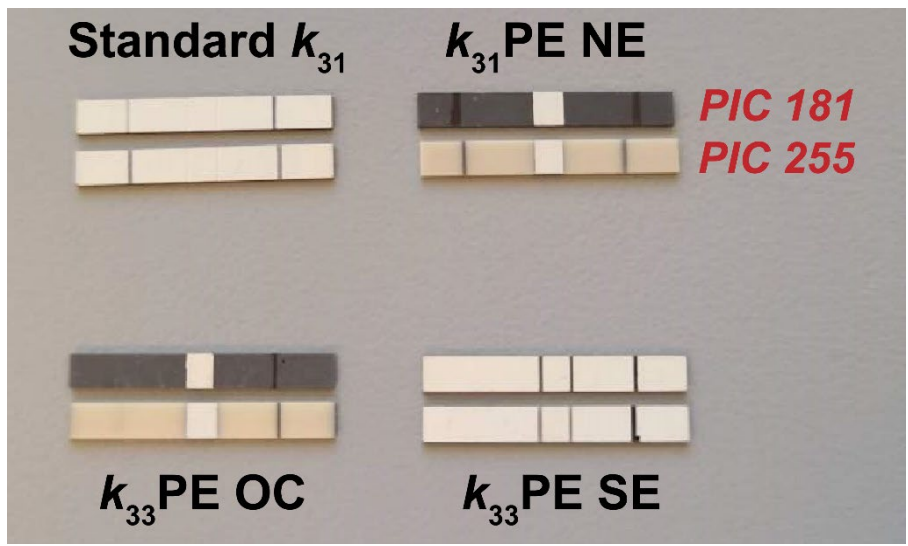

**Figure 3.** Actual piezoelectric samples used in this study. Black lines denote polarization direction; faces with 2 black lines represent positive side (arrowhead) of polarization. See Figure 1 and 2 for specific geometry of each sample.

## 3. Theory/Calculation

*3.1 Parameter Determination Using Standard $k_{31}$ Mode*

Determination of physical parameters of Standard $k_{31}$ specimen was already shown in IEEE Standard and Zhuang *et al.*[24,30]. The basic formulation starts from admittance equation, which is given by:

$$Y_{31}^* = j\omega \frac{\varepsilon_0 \varepsilon_{33}^{X*} wl}{t} \left[ (1 - k_{31}^{*2}) + k_{31}^{*2} \frac{tan(\omega l/2v_{11}^{E*})}{\omega l/2v_{11}^{E*}} \right] \quad (7)$$

Where $\varepsilon_{33}^{X*}$ is complex intensive (stress-free) relative permittivity, $v_{11}^{E*}$ is complex sound velocity, which is defined as $v_{11}^{E*} = 1/\sqrt{\rho s_{11}^{E*}}$ with mass density $\rho$ and complex intensive (E-constant) elastic compliance $s_{11}^{E*}$, and $k_{31}^*$ is complex electromechanical coupling coefficient, which is defined as $k_{31}^{*2} = d_{31}^{*2}/(\varepsilon_0 \varepsilon_{33}^{X*} s_{11}^{E*})$, with complex piezoelectric coefficient $d_{31}^*$.

$\varepsilon_{33}^X$ and corresponding dielectric loss ($\tan \delta_{33}'$) can be determined from off-resonance capacitance and phase lag measurement, respectively. $s_{11}^E$ and corresponding elastic loss ($\tan \phi_{11}'$) can be determined with the following equations from admittance resonance spectrum:

$$s_{11}^E = \frac{1}{4\rho f_A^2 l^2} \quad (8)$$

$$\tan \phi_{11}' = \frac{1}{Q_A} \quad (9)$$

Resonance ($f_A$) and antiresonance frequency ($f_B$) are maximum admittance and impedance point, respectively. $Q_A$ and $Q_B$, which are corresponding mechanical quality factors, can be determined by 3 dB method[17] or quadrantal bandwidth method[27]. Electromechanical coupling factor, $k_{31}$, is determined by the following equation:

$$\frac{k_{31}^2}{1 - k_{31}^2} = \frac{\pi}{2} \frac{f_B}{f_A} \tan\left(\frac{f_B - f_A}{f_A}\right) \quad (10)$$

After $\varepsilon_{33}^X$, $s_{11}^E$ and $k_{31}$ are obtained, $d_{31}$ can be obtained using the following equation:

$$d_{31}^2 = k_{31}^2 (\varepsilon_0 \varepsilon_{33}^X s_{11}^E) \quad (11)$$

Finally, intensive piezoelectric loss ($\tan \theta_{31}'$) can be determined by using the following equation:

$$\tan \theta'_{31} = \frac{\tan \delta'_{33} + \tan \phi'_{11}}{2} + \frac{1}{4}\left(\frac{1}{Q_A} - \frac{1}{Q_B}\right)\left[1 + \left(\frac{1}{k_{31}} - k_{31}\right)^2 \Omega_B^2\right] \quad (12)$$

Where $k_{31}$ is real part of complex electromechanical coupling factor and $\Omega_B$ is normalized antiresonance frequency, which is represented in terms of antiresonance angular frequency defined by $\omega_B = 2\pi f_B$ and $s_{11}^E$-related sound velocity ($v_{11}^E$):

$$\Omega_B = \frac{\omega_B l}{2 v_{11}^E} \quad (13)$$

The electromechanical coupling square loss, imaginary part of $k_{31}^{*2}$, can be determined by the following equation:

$$\tan \chi_{31} = 2\tan \theta'_{31} - \tan \delta'_{33} - \tan \phi'_{11} \quad (14)$$

With Standard $k_{31}$ specimen, the only real extensive parameter that can be obtained is extensive elastic compliance ($s_{11}^D$), which is defined as:

$$s_{11}^D = s_{11}^E (1 - k_{31}^2) \quad (15)$$

Extensive (strain-free) dielectric permittivity ($\varepsilon_{33}^x$) can be obtained from the damped capacitance with thickness-mode ($k_t$) plate, by measuring admittance spectrum around the fundamental resonance frequency. However, since the measurement accuracy is low in high frequency region, especially in MHz regime, researchers indirectly calculate $\varepsilon_{33}^x$ by using complicated equations originated from piezoelectric constitutive relations[31].

All the loss parameters, which can be determined from the equations in section 3.1, are intensive losses. In order to obtain extensive loss, one must utilize the following matrix equation[17]:

$$\begin{bmatrix} \tan \delta' \\ \tan \phi' \\ \tan \theta' \end{bmatrix} = [K] \begin{bmatrix} \tan \delta \\ \tan \phi \\ \tan \theta \end{bmatrix} \quad (16)$$

Where $[K]$ is 3×3 matrix that is called $[K]$ matrix, which is defined as:

$$[K] = \begin{bmatrix} 1 & k^2 & -2k^2 \\ k^2 & 1 & -2k^2 \\ 1 & 1 & -1-k^2 \end{bmatrix} \quad (17)$$

$[K]$ is involutory ($[K] = [K]^{-1}$), so that inverse relationship of Equation (16) should also hold. It is noteworthy to mention that the $[K]$ matrix relationship Equation (16) is valid only for $k_{31}$ type, not for $k_{33}$ type. Pure extensive loss for $k_{33}$ mode type can rather be obtained in 3D constrained conditions.

So far, physical parameter determination process of both intensive and extensive parameters for the Standard $k_{31}$ mode has been discussed. It is noteworthy to mention that obtaining extensive parameters requires additional steps. Furthermore, from $k$-matrix formulation, it should be noted that the errors for extensive losses become larger due to error propagation.

*3.2 Parameter Determination Using Standard $k_{33}$ Mode*

For $k_{33}$ mode, the admittance equation is given by:

$$Y_{33}^* = \frac{j\omega wt \varepsilon_0 \varepsilon_{33}^{X*}(1-k_{33}^{*2})}{l\left[1 - k_{33}^{*2}\frac{tan(\omega l/2v_{33}^{D*})}{\omega l/2v_{33}^{D*}}\right]} \quad (18)$$

where $k_{33}^{*2}$ is electromechanical coupling factor defined as $k_{33}^{*2} = d_{33}^{*2}/(\varepsilon_0\varepsilon_{33}^{X*}s_{33}^{E*})$ and $v_{33}^{D*}$ is extensive elastic compliance ($s_{33}^{D*}$)-related sound velocity; dimension is defined in Figure 1 (b). $\varepsilon_{33}^X$ can also be measured with Standard $k_{33}$ specimens; however, because of intrinsic geometry, $\varepsilon_{33}^X$ measured from $k_{33}$ specimens are normally overestimated depending on the sample's aspect ratio[18,25,26,32]. Therefore, to obtain $\varepsilon_{33}^X$, either Standard $k_{31}$ specimens or thickness mode plates should be utilized.

The admittance equations of $k_{31}$ mode and $k_{33}$ mode have different resonance conditions, as well as different electrical boundary conditions. For instance, $k_{31}$ mode has *E*- constant condition because of free charges due to surface electrode that cancels out depolarization field, whereas $k_{33}$ mode, which has $D$-constant condition, does not have any free charges that cancel out the depolarization field [10]. Due to different electrical boundary conditions that affects elasticity, they also have different resonance conditions: From Eq. (7), $Y_{31}^* = \infty$ when $tan(\omega l/2v_{11}^{E*}) = \infty$, whereas Eq. (17) gives $Y_{33}^* = 0$ when $tan(\omega l/2v_{33}^{D*}) = \infty$. Therefore, different from $k_{31}$ mode, in which elastic compliance is obtained from resonance frequency, $k_{33}$ mode has a half-wave resonating condition at the antiresonance frequency, and $s_{33}^D$ is obtained with the following equation:

$$s_{33}^D = \frac{1}{4\rho f_B^2 l^2} \tag{19}$$

The complex elastic constants of $k_{33}$ mode ($s_{33}^{D*}$) is not perfectly $D$-constant, due to mechanical boundary condition. Assuming 3 is direction along $l$ and 1 and 2 are along $t$ and $w$, respectively (see Figure 1 (b)), and $l$ is much larger than $w$ and $t$, the condition $X_1 = X_2 = 0$ satisfies, meaning that there is no depolarization field along 1 and 2 directions. Since it is the case, $k_{33}$ mode specimen has $E$-constant in those directions. In 3D notation proposed by Ikeda [27], the elastic compliance is given by $s_{33}^{EED*}$, which means $E_1$, $E_2$ and $D_3$ are constant. Therefore, the imaginary part of $s_{33}^{D*}$ is represented as triple-prime loss ($\tan \phi_{33}'''$), to be distinguished from purely extensive loss ($\tan \phi_{33}$). $\tan \phi_{33}'''$ is obtained from $Q_B$ of $k_{33}$ mode and defined by the following expression:

$$\tan \phi_{33}''' = \frac{1}{Q_B} = \frac{1}{1-k_{33}^2}[\tan \phi_{33}' - k_{33}^2(2 \tan \theta_{33}' - \tan \delta_{33}')] \tag{20}$$

Note that Equation (20) follows the [K] matrix, but the relationship is between triple-prime and single-prime, not between non-prime and single-prime losses. Similar to $k_{31}$ mode, electromechanical coupling factor for $k_{33}$ mode can be determined with $f_A$ and $f_B$:

$$k_{33}^2 = \frac{\pi}{2}\frac{f_A}{f_B}\tan\left(\frac{f_B - f_A}{f_B}\right) \qquad (21)$$

Different from $k_{31}$ mode, intensive elastic compliance ($s_{33}^E$) can be indirectly obtained from the following equation:

$$s_{33}^E = \frac{s_{33}^D}{(1 - k_{33}^2)} \qquad (22)$$

In order to obtain intensive elastic loss ($\tan \phi'_{33}$) and intensive piezoelectric loss ($\tan \theta'_{33}$), the following two equations should be utilized:

$$\tan \phi'_{33} - 2k_{33}^2 \tan \theta'_{33} = \frac{(1 - k_{33}^2)}{Q_{B,33}} - k_{33}^2 \tan \delta'_{33} \qquad (23)$$

$$\tan \phi'_{33} + 2\tan \theta'_{33} \qquad (24)$$
$$= -\left(\frac{1}{Q_{A,33}} - \frac{1}{Q_{B,33}}\right)\frac{(k_{33}^2 - 1 + \Omega_{A,33}^2/k_{33}^2)}{2} - \tan \delta'_{33}$$

$\tan \phi_{33}$ and extensive piezoelectric loss ($\tan \theta_{33}$) cannot be obtained directly with $k_{33}$ mode specimen, but rather can be obtained by measuring $k_t$ mode plate. Since $k_t$ mode plates have nonzero stress along 1 and 2 directions (two orthogonal directions that are both perpendicular to polarization), there exist depolarization field along those directions; therefore, with 3D notation, the elastic stiffness is $c_{33}^{DDD*}$ and the corresponding loss is $\tan \phi_{33}$.

By now, the method to obtain different real and imaginary parameters using Standard $k_{31}$ and $k_{33}$ mode has been discussed. In usual case, the parameters that are not directly determined from resonance frequencies and quality factors, but rather determined from the equations with other physical parameters, have larger statistical errors due to error propagation process. To the extreme, Zhuang[31] reported 100 % statistical error on $\tan \theta_{33}$ on a soft PZT.

*3.3 Analytical Admittance Equation of PE configuration*

The derivation process for analytical admittance equation of the PE configuration has already been discussed in our previous papers [29]. If the admittance equation is simplified, the universal admittance equation is given by:

$$Y_{PE} = jw \left[ \frac{2d_{31}^{*2} v_{11}^{E*} v_{side}^{*} s_{side}^{*}}{ts_{11}^{E*} \left[ \frac{s_{side}^{*} v_{side}^{*}}{\tan\left(\frac{a\omega l}{2v_{11}^{E*}}\right)} - v_{11}^{E*} s_{11}^{E*} \tan\left(\frac{(1-a)\omega l}{2v_{side}^{*}}\right) \right]} + \frac{a l \omega \varepsilon_0 \varepsilon_{33}^{X*}(1 - k_{31}^{*2})}{t} \right] \quad (25)$$

Where $a$ is portion of center part ($0 < a < 1$), $s_{side}^{*}$ is elastic compliance of side part along the length and $v_{side}^{*}$ is sound velocity along the length of side part. Depending on the electrical boundary or poling conditions of side part, $s_{side}^{*}$ can be $s_{11}^{D*}$ ($k_{31}$ NE), $s_{33}^{D*}$ ($k_{33}$ OC) or $s_{33}^{E*}$ ($k_{33}$ SE), and the corresponding elastic loss (imaginary part) values in the complex notation are $\tan\phi_{11}$, $\tan\phi_{33}'''$, and $\tan\phi_{33}'$, respectively. In Equation (25), the first term involving tangent functions is combined motional admittance, and the second term is damped admittance. The damped admittance is identical to that of $k_{31}$ mode specimen in Equation (7), with different length ($al$, because only center portion is considered).

For motional admittance part in Equation (25), taking out $\frac{d_{31}^{*2}}{s_{11}^{E*2}}$ term outside the square bracket, then making numerator 1 by dividing both numerator and denominator by numerator, and using the fact that $v \times s = \frac{1}{\rho v}$, we obtain:

$$Y_{m,PE} = \frac{jw}{t} \frac{2d_{31}^{*2}}{s_{11}^{E*2}} \left[ \frac{1}{\left[ \rho v_{11}^{E*} \cot\left(\frac{a\omega l}{2v_{11}^{E*}}\right) - \rho v_{side}^{*} \tan\left(\frac{(1-a)\omega l}{2v_{side}^{*}}\right) \right]} \right] \quad (26)$$

For simplicity, it is useful to discuss impedance form of Equation (26), which is:

$$Z_{m,PE} = \frac{-jt}{2w} \frac{s_{11}^{E*2}}{d_{31}^{*2}} \left[ \rho v_{11}^{E*} \cot\left(\frac{a\omega l}{2v_{11}^{E*}}\right) - \rho v_{side}^{*} \tan\left(\frac{(1-a)\omega l}{2v_{side}^{*}}\right) \right] \quad (27)$$

Equation (27) shows clear separation of motional impedance for center and side part. Including multiplication term outside the square bracket, the first term inside the square bracket with cotangent function is impedance of the center part and the second term with tangent function is impedance of the side part. To confirm the equation, $a = 1$ can be put into the equation. In the case of $a = 1$, the center part takes 100 % of the entire geometry, so the equation describes Standard $k_{31}$ mode, and Equation (25) turns into Equation (7). In the case of $a$ approaches to zero, resonance frequencies and mechanical quality factors in an admittance curve approaches to reflect elasticity of side part.

*3.4 Parameter Determination Process Using Standard $k_{31}$ and PE Specimens*

The following steps are parameter determination process using Standard $k_{31}$ mode samples and PE samples:

1. The dimensions ($l$, $w$, $t$), mass ($\rho$) and the center portion ($a$) for PE should be measured for each sample. This procedure is always required for any piezoelectric specimen measurement.

2. Admittance curves of Standard $k_{31}$ mode samples should be measured, and related parameters ($s_{11}^{E*}, \varepsilon_{33}^{X}, d_{33}^{*}, k_{31}^{*}$) are obtained by using the equations shown in 3.1.1. Unlike Standard $k_{33}$ mode specimens that have several issues, Standard $k_{31}$ mode specimens do not have particular issues.

3. Admittance curves of PE specimens ($k_{31}$ NE, $k_{33}$ OC and $k_{33}$ SE) should be measured, and elastic parameters are obtained with nonlinear regression curve fitting. Since the center portion of the PE specimens is $k_{31}$ mode, $k_{31}$ mode-related parameters exist in analytical solutions of PE configuration, as the form of damped and motional admittance. In order to minimize the fitting variables, intensive parameters determined from $k_{31}$ mode in step 2 can be plugged into analytical solution when the fitting is performed. Therefore, for each PE specimen, there are only 2 parameters (elastic compliance and corresponding elastic loss of side part) that are needed to be determined. It is noteworthy to mention that Majzoubi et al.[17] proved 1D assumption $s_{11}^{D*} \approx 1/c_{11}^{D*}$ holds; therefore, the imaginary part of $s_{11}^{D*}$ is purely extensive loss ($\tan \phi_{11}$).

4. $k_{33}$ and $d_{33}$ can be determined from the following equations, after $s_{33}^D$ and $s_{33}^E$ are determined from $k_{33}$ OC and $k_{33}$ SE, respectively:

$$k_{33} = \sqrt{1 - \frac{s_{33}^D}{s_{33}^E}} \tag{28}$$

$$d_{33} = k_{33}\sqrt{\varepsilon_0 \varepsilon_{33}^X s_{33}^E} \tag{29}$$

These parameters, according to IEEE Standard[24], are determined from Standard $k_{33}$ mode specimens. However, with Standard $k_{33}$ mode specimens, reliable data cannot be obtained, because there are several issues, such as high impedance values near antiresonance frequency that causes huge noise in experimental admittance/impedance curves, indispensable wire attachment that shifts antiresonance frequency, and fringing electric field issue due to intrinsic geometry. Therefore, With the aid of PE specimens, these parameters can be obtained more reliably.

5. Finally, $\tan \theta'_{33}$ and $\tan \chi_{33}$ can be determined using the following equation[31]:

$$\tan \theta'_{33} = \frac{1}{2k_{33}^2}[\tan \phi'_{33} - (1 - k_{33}^2)\tan \phi'''_{33} + k_{33}^2 \tan \delta'_{33}] \tag{30}$$

$$\tan \chi_{33} = 2 \tan \theta'_{33} - \tan \delta'_{33} - \tan \phi'_{33} \tag{31}$$

So far, determination method of all the intensive parameters and extensive-like elastic parameters of $k_{31}$ and $k_{33}$ mode has been discussed, with combination of Standard $k_{31}$ mode and three PE samples. As already mentioned in section 3.2, other extensive parameters cannot solely be obtained from $k_{31}$ and $k_{33}$ mode but obtained by measuring $k_t$ mode samples. Figure

4 summarizes parameter determination process using Standard $k_{31}$ mode and three PE samples.

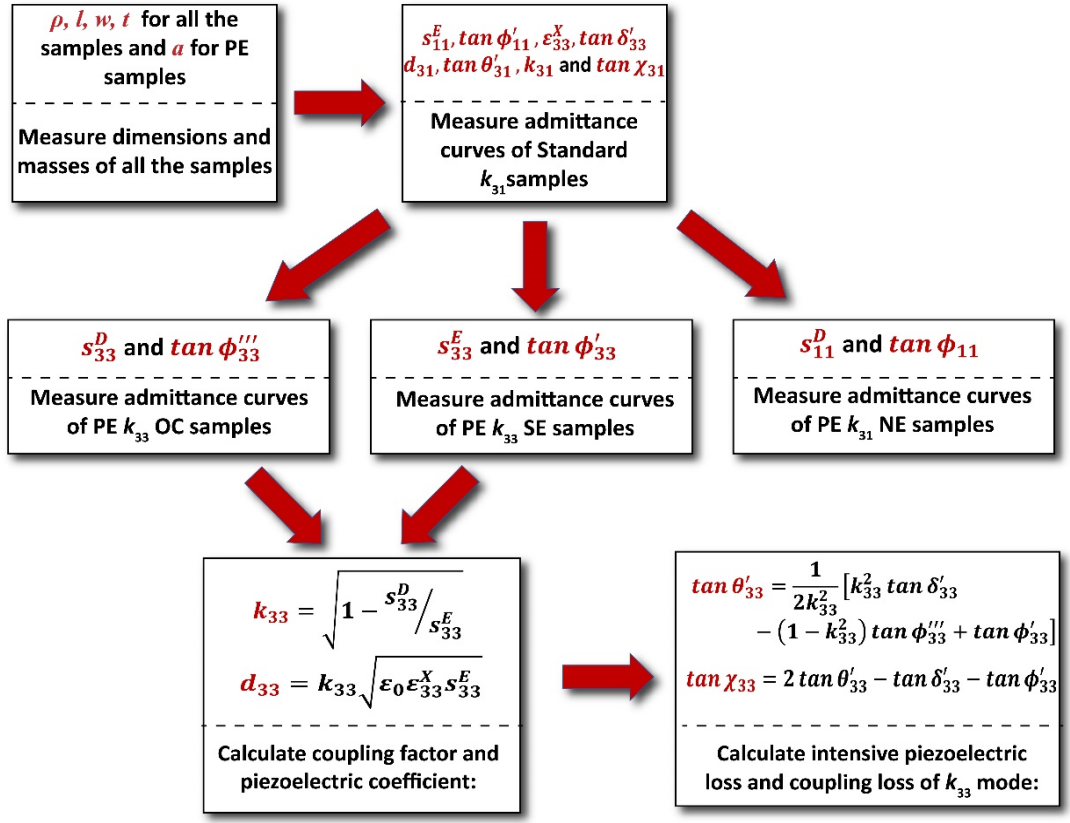

**Figure 4.** Parameter determination process using Standard $k_{31}$ and three types of PE samples.

### 3.5 Simplification of PE Configuration

In both Majzoubi *et al.*[17] and our previous work[18], both $k_{31}$ and $k_{33}$ PE configurations included open circuit (OC) and short circuit (SC). In the case of $k_{31}$ mode PE, this inclusion was necessary to characterize both resonance and antiresonance frequencies and corresponding mechanical quality factors; in the case of $k_{33}$ mode PE, SC specimen was necessary to obtain $d_{33}$ and $\tan\theta'_{33}$. However, in this study, some of the PE geometries are omitted for the following reasons:

1. Since Standard $k_{31}$ mode does not have particular issues, OC (for antiresonance characterization) and SC (for resonance characterization) of $k_{31}$ PE are not needed. Including two geometry brings out redundant tasks. Only geometry needed for $k_{31}$

PE is NE, since it allows direct determination of extensive elastic compliance and extensive elastic loss.

2. For $k_{33}$ mode PE, SC was omitted, because of complicated process to obtain $d_{33}$ and $\tan \theta'_{33}$. In order to obtain these parameters using SC specimen, $s_{33}^{D*}$ and $s_{33}^{E*}$ should be determined first and plugged into very complicated analytical admittance equation[18,29]. Furthermore, attaching wires to SC specimens possibly distort experimental admittance/impedance curves, as soldering iron and wires can add the mass to the specimens. Without using SC specimens, $d_{33}$ and $\tan \theta'_{33}$ can be obtained by using Equation (29) and (30), after determination of $s_{33}^{D*}$ and $s_{33}^{E*}$ from $k_{33}$ PE OC and $k_{33}$ PE SE, respectively.

3. Reducing types of samples would greatly accelerate sample preparation, measurement, and analysis process.

Therefore, with omitted OC and SC for $k_{31}$ mode PE and SC for $k_{33}$ mode PE, the modified PE method includes NE for $k_{31}$ mode PE, OC and SE for $k_{33}$ mode PE, along with Standard $k_{31}$ specimen.

## 4. Results

### 4.1 Possible Error Causes of PE Specimens

In this section, the factors, some of which are demonstrated with FEA simulation results, that possibly causes the error for PE specimens will be discussed. Those factors can be fringing electric field at center part, and partial poling issue near the boundary between center and side part.

#### 4.1.1 Fringing Electric Field Occurring at the Center Part

The analytical solutions for PE configurations were already verified with FEA in previous works[29]. Previously, the differences between admittance equations of analytical solutions and FEA were discussed: the height (baseline) difference and peak values[29]. With basic intuition, it is noticed that the center portion must experience fringing electric field, due to the fact that it is surrounded with side part that has similar magnitude of dielectric permittivity. In order to profoundly investigate what may cause the differences, FEA simulation was performed. Table 1 shows the input $\varepsilon_{33}^X$ used in the simulation, along with $\varepsilon_{33}^X$ determined from center portion's impedance value generated by simulation of each PE. Somehow, in the case of $k_{33}$ PE SE, the electrode on the side part may suppress the fringing electric field, so almost no change occurs in $\varepsilon_{33}^X$. However, in the case of both $k_{31}$ PE NE and $k_{33}$ PE OC, significant overestimation of $\varepsilon_{33}^X$ is observed, though smaller value was used as input for the simulation. We also reported overestimation of $\varepsilon_{33}^X$ obtained from center part of PE samples[16]. Therefore, in order to minimize the difference, the overestimated values of $\varepsilon_{33}^X$ is used for analytical solutions, rather than the same input FEA parameter, in order to calibrate the overestimation of $\varepsilon_{33}^X$ at the center portion. Figure 5 shows analytical and FEA admittance curves of $k_{31}$ PE NE and $k_{33}$ PE OC, with and without permittivity calibration. Without calibration, the height (baseline) difference is obvious, whereas the height of admittance curves with the calibration shows much better agreement between analytical solutions and FEA. Since degree of overestimation is different from each sample due to different $a$, $\varepsilon_{33}^X$ should be directly measured from the center part of each sample and used as fitting parameter, rather than using $\varepsilon_{33}^X$ determined from Standard $k_{31}$ plate. The remaining small difference in peak values may be due to the difference between 1D consideration of analytical solution and 3D consideration of FEA. Despite the differences in peak values, less than 0.3 % difference occurs for resonance frequencies, and less than 1.6 % difference occurred for quality factors.

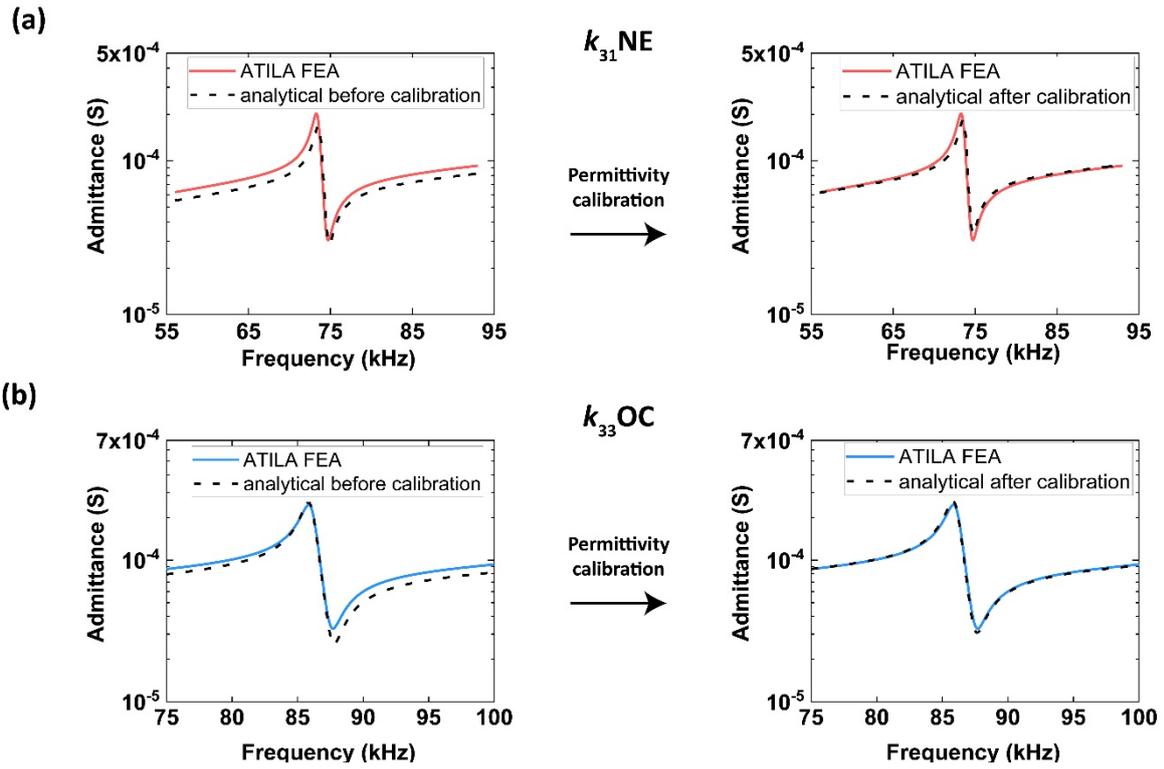

**Figure 5.** Admittance curves generated from ATILA FEA, and analytical solution with and without permittivity calibration of (a)PE $k_{31}$ NE and (b) PE $k_{33}$ OC.

$\varepsilon_{33}^X$ values from ATILA FEA

| Input | Output from $k_{31}$ PE NE center | Output from $k_{33}$ PE OC center | Output from $k_{33}$ PE SE center |
|---|---|---|---|
| 1700 | 1872 | 1918 | 1707 |

**Table 1.** $\varepsilon_{33}^X$ used as input the simulation (PZT 5A), along with $\varepsilon_{33}^X$ determined from center portion's impedance values generated by simulation of each PE for comparison.

4.1.2 Partial Poling Issues in $k_{33}$ PE Specimens

For $k_{33}$ PE OC and $k_{33}$ PE SE, two-step poling process is involved: bulk ceramic is poled, then cut into thin plates with desired sample dimension, center part of each piece is electroded and re-poled. The process is required, since the side part should have polarization along the length direction for $k_{33}$ mode elastic characterization, whereas the center actuation part should always be $k_{31}$ mode, having polarization along thickness direction. However, in the process of two-step poling with two different directions, canted poling may occur at the boundary between center and side part. Since the part with canted polarization has different physical properties compared to upward and side polarization, it would affect experimental admittance curve.

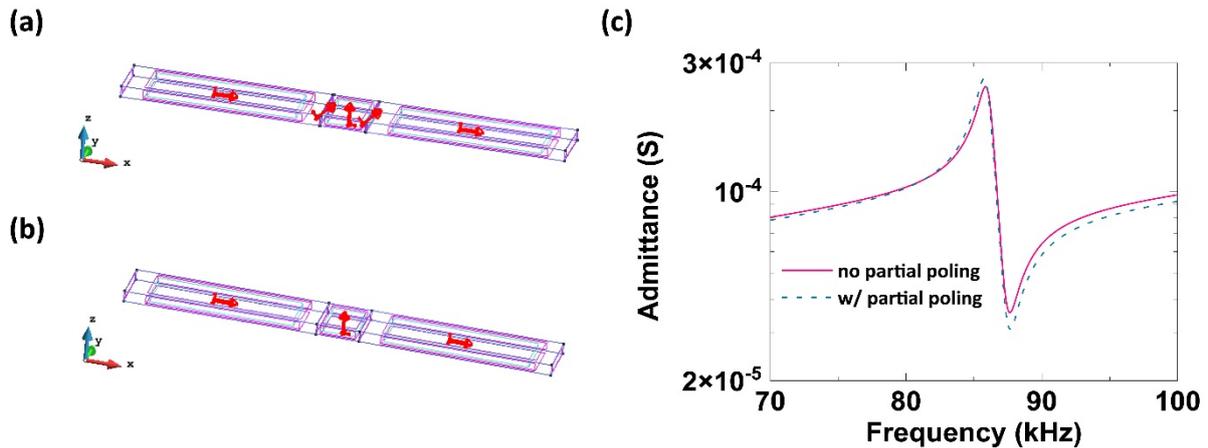

**Figure 6.** ATILA FEA geometry for (a) PE with partial poling (45° at the interface of side and center) and (b) without partial poling. (c) Admittance curve difference between two cases.

In order to investigate the effect of canted polarization, ATILA FEA simulation was performed. Two volumes, each of which has 50 μm gap in the length direction with 45° canted polarization, were located at each end of the center portion in PE geometry, as shown in Figure 6 (a) and (b). Figure 6 (c) shows the comparison of admittance spectra of two cases: PE geometry with and without the volumes with canted polarization. Even though volumes with canted polarization were inserted in both edge of the center portion, the admittance curve does not show significant difference, when compared to the admittance curves of PE without the volumes with canted polarization. In order to make quantitative comparison, $f_A$, $f_B$, $Q_A$ and $Q_B$ of two admittance curves are compared. Table 2 shows the values of these parameters, along with the percentage differences. The percentage difference in frequency values ($f_A$ and $f_B$) ranges from 0.1 % to 0.2 %, and the difference in quality factors ($Q_A$ and $Q_B$) ranges from 0.26 – 0.4 %. Therefore, partial canted polarization that may be occurred at the boundary between center and side part does not significantly affect measured admittance curves, when considering FEA simulation results.

|  | $f_a$ (KHz) | $Q_a$ … | $f_b$ (KHz) | $Q_b$ … |
|---|---|---|---|---|
| w/o partial poling | 86.1 | 76.9 | 87.4 | 83.7 |
| w/ partial poling | 85.9 | 76.7 | 87.5 | 84.1 |
| Difference (%) | 0.2 % | 0.26 % | 0.1 % | 0.4 % |

**Table 2.** values of $f_A$, $f_B$, $Q_A$ and $Q_B$ and percentage difference in the cases of with and without partial poling volumes in $k_{33}$ PE OC with input parameters of PZT 5A.

*4.2 Parameter Determination*

Table 3 shows parameters determined from PE fitting method. The real and imaginary parameters determined from $k_{31}$ mode samples (See supplementary materials for $k_{31}$ mode parameters) were used to minimize fitting variables, and permittivity calibration discussed in section 4.1.1 was also applied. The fitting curves, along with experimental admittance curves, near resonance and antiresonance peaks are shown in Figure 7 (Fittings of full curves are shown in supplementary materials). All the percentage fitting errors were less than 1 %, which represents great fit of analytical solutions to experimental admittance curves. In table 5, in terms of statistical variation, PE samples show similar error range, compared to Standard $k_{31}$ samples. Therefore, PE method is as reliable as standard method. The values in Table 3 have slight discrepancies when compared to physical parameter values of PIC 255 in our previous report[18]. This may be due to that they were made from different ceramic blocks, as well as difference in electrode materials.

**Parameters determined from PE samples**

| PIC 181 | | | | |
|---|---|---|---|---|
| Real Parameters | | | | |
| $s_{11}^D$ ($\times 10^{-12}$ m$^2$/N) | $s_{33}^D$ ($\times 10^{-12}$ m$^2$/N) | $s_{33}^E$ ($\times 10^{-12}$ m$^2$/N) | $d_{33}$ (pC/N) | $k_{33}$ …. |
| 10.53 $\pm$ 0.05 | 8.53 $\pm$ 0.04 | 13.03 $\pm$ 0.06 | 224 $\pm$ 2 | 0.588 $\pm$ 0.004 |
| Imaginary Parameters | | | | |
| $\tan \phi_{11}$ (%) | $\tan \phi_{33}'''$ (%) | $\tan \phi_{33}'$ (%) | $\tan \theta_{33}'$ (%) | $\tan \chi_{33}$ (%) |
| 0.039 $\pm$ 0.001 | 0.030 $\pm$ 0.001 | 0.053 $\pm$ 0.002 | 0.229 $\pm$ 0.003 | 0.043 $\pm$ 0.007 |
| PIC 255 | | | | |
| Real Parameters | | | | |
| $s_{11}^D$ ($\times 10^{-12}$ m$^2$/N) | $s_{33}^D$ ($\times 10^{-12}$ m$^2$/N) | $s_{33}^E$ ($\times 10^{-12}$ m$^2$/N) | $d_{33}$ (pC/N) | $k_{33}$ …. |
| 14.30 $\pm$ 0.05 | 9.68 $\pm$ 0.03 | 17.48 $\pm$ 0.1 | 365 $\pm$ 5 | 0.668 $\pm$ 0.001 |
| Imaginary Parameters | | | | |
| $\tan \phi_{11}$ (%) | $\tan \phi_{33}'''$ (%) | $\tan \phi_{33}'$ (%) | $\tan \theta_{33}'$ (%) | $\tan \chi_{33}$ (%) |
| 0.87 $\pm$ 0.01 | 0.51 $\pm$ 0.01 | 1.19 $\pm$ 0.01 | 1.79 $\pm$ 0.02 | 0.85 $\pm$ 0.04 |

**Table 3.** Real and imaginary parameters of PIC 181 and PIC 255 determined from PE method. Errors are from data variation of 6 samples.

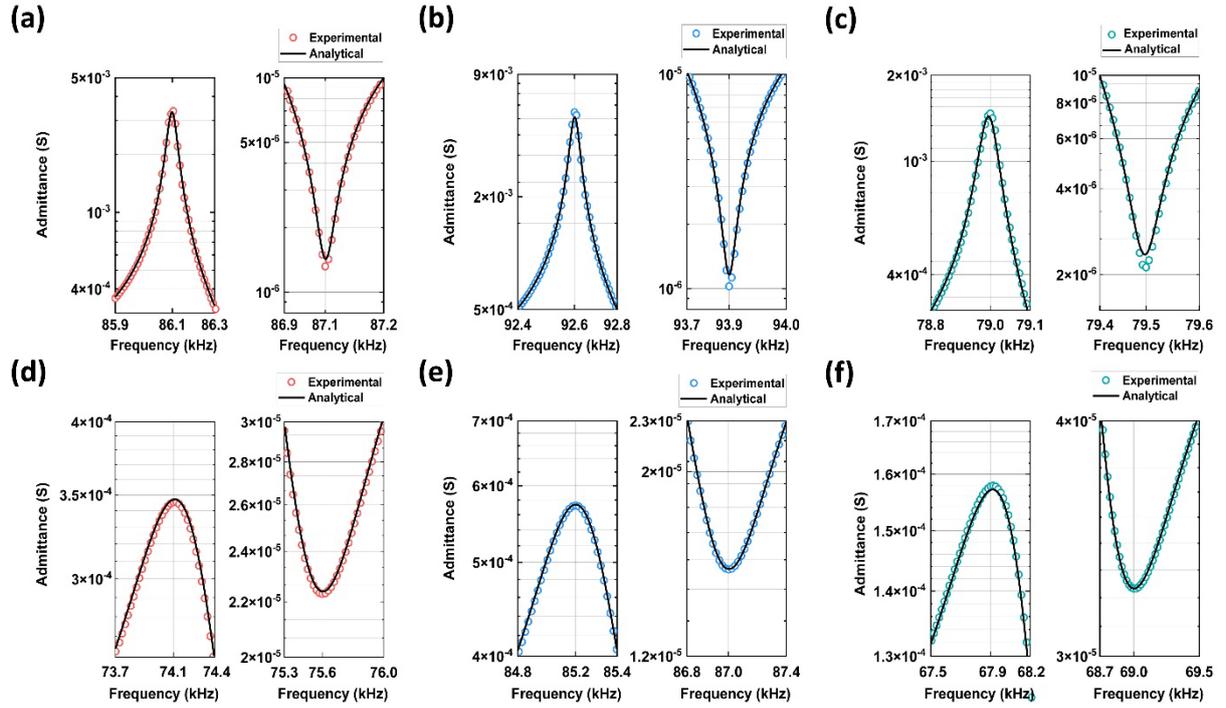

**Figure 7.** Experimental admittance curves of PE samples made of PIC 181 and PIC 255 measured with impedance analyzer. Black lines are analytical fitting curves.

## 5. Discussion

### 5.1 *Benefits of PE Configuration as a Tool for Parameter Characterization*

#### 5.1.1 Tool for Extensive Elastic Loss Characterization for $k_{31}$ mode

As aforementioned, Standard $k_{31}$ mode does not have issues when being utilized for parameter determination. PE configuration ($k_{31}$ NE) was proposed because it allows direct determination of extensive elastic compliance ($s_{11}^D$) and extensive elastic loss ($\tan\phi_{11}$). In order to obtain $s_{11}^D$ and $\tan\phi_{11}$ with only Standard $k_{31}$ mode specimen, Equation (15), (16) and (17) must be applied. Though error propagation process may not exaggerate the error of $s_{11}^D$ so significantly, the error for $\tan\phi_{11}$ becomes enlarged, due to complicated [K] matrix

equation. When expanded, the $k$-matrix provides the following equation for $\tan\phi_{11}$:

$$\tan\phi_{11} = \frac{1}{1-k_{31}^2}(k_{31}^2 \tan\delta'_{33} + \tan\phi'_{11} - 2k_{31}^2 \tan\theta'_{31}) \tag{32}$$

With Equation (32), it is noteworthy to mention that, if $\tan\phi_{11}$ is obtained with Standard $k_{31}$ mode, the error for $k_{31}$, $\tan\delta'_{33}$, $\tan\phi'_{11}$, and $\tan\theta'_{31}$ are all accumulated to $\tan\phi_{11}$. PE method proposed in this study, on the other hand, provides $\tan\phi_{11}$ (from curve fitting) that has an error comparable with $\tan\phi'_{11}$, which is directly determined from $Q_A$ of Standard $k_{31}$ mode specimen.

### 5.1.2 PE as Substitution for Standard $k_{33}$ Specimen

Different from standard $k_{31}$ specimen, standard $k_{33}$ specimen has several issues that hinders researchers from obtaining accurate physical parameters. The most significant problem of standard $k_{33}$ specimen is high impedance value near antiresonance frequency and corresponding 3 dB bandwidth.

Most impedance analyzers have accuracy limit near $10^7 - 10^8$ Ω; above near $10^8$ Ω, the measurement error becomes larger than 10 %[33-36]. Among 4 impedance analyzers that we investigated, Agilent 4294A Precision Impedance Analyzer has the lowest measurement error in the frequency regime from 1 KHz to 1 MHz, which corresponds to fundamental frequencies of most millimeter-scale IEEE Standard $k_{31}$ and $k_{33}$ samples. Figure 8 illustrates measurement error of Agilent 4294A Precision Impedance Analyzer, in terms of impedance and sweeping frequency[33]. As seen from Figure 8, from 100 Hz to 200 KHz, the measurement error exceeds 10 % at $4\times10^7$ Ω. For the case of $k_{33}$ mode specimen made from soft PZT, We previously reported impedance value about $2 \times 10^8$ Ω near antiresonance frequency and corresponding 3 dB bandwidth, along with large fluctuation (electrical noise) of impedance values[18]. Hard PZT, which has much larger mechanical quality factors than soft

PZT due to domain wall pinning, is likely to suffer more on low measurement accuracy near antiresonance frequency.

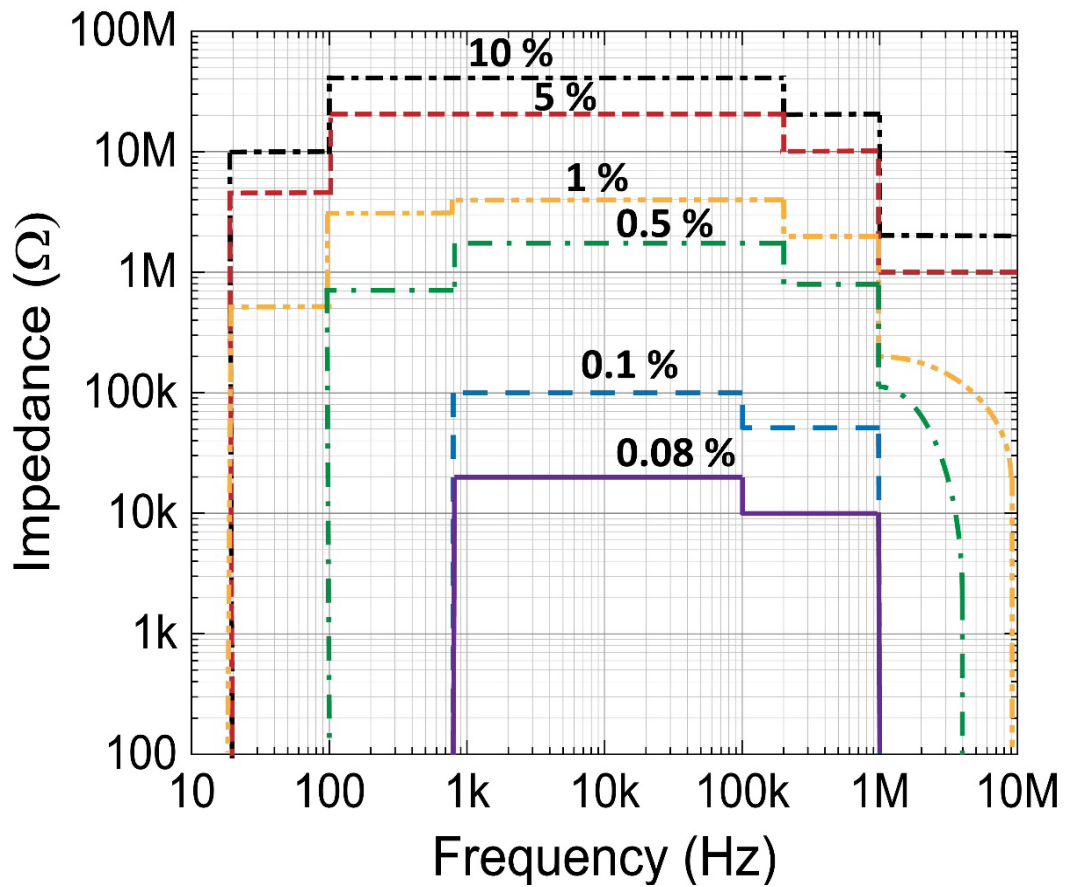

**Figure 8.** Measurement error in terms of impedance (in Ω) and operating frequency (in Hz) of Agilent 4294A Precision Impedance Analyzer. Redrawn from [29].

In order to see feasibility of impedance analyzer on $k_{33}$ mode specimens made of various, and commonly utilized piezoelectric materials, FEA simulation was performed near antiresonance and corresponding 3 dB bandwidth; impedance values at antiresonance frequency and near 3 dB bandwidth for PZT 5A (soft), PZT 4 (semi-hard) and PZT 8 (hard) are shown in Table 4. For $k_{33}$ mode geometry used for the simulation, dimension of 5 × 5 × 20 mm (1 to 4 aspect ratio) was utilized, since the width to length aspect ratio of $k_{33}$ mode specimens adapted by researchers range from 1:3 to 1:5[37-39]. The impedance values of PZT 5A near $f_B$ falls into measurement error range from 1 % to 5 %, which can be considered as okay values. However, those of PZT 4 fall into region of error more than 10 %, and those of PZT 8 are totally out of scope. Therefore, using $k_{33}$ mode geometry for measurement with impedance analyzer is not appropriate.

|          | Impedance at $f_B$ (Ω) | Impedance at 3 dB of $f_B$ (Ω) |
|----------|------------------------|-------------------------------|
| PZT 5A   | 7.35×10$^6$            | 5.20×10$^6$                   |
| PZT 4    | 7.15×10$^7$            | 5.05×10$^7$                   |
| PZT 8    | 2.07×10$^8$            | 1.46×10$^8$                   |

**Table 4.** Impedance values at antiresonance frequency and 3 dB bandwidth for PZT 5A (soft), PZT 4 (semi-hard) and PZT 8 (hard), calculated from FEA simulation.

In terms of impedance analyzer's measurement accuracy, PE, on the other hand, provides much more reliable admittance/impedance values than Standard $k_{33}$ mode specimen. It is noteworthy to mention that PIC 181, hard PZT used in this study, has $Q_m \sim 2000$, which is much larger than that of PZT 8 ($Q_m \sim 1000$). According to FEA results shown in Table 4 and Figure 8, even the impedance values near $f_B$ and corresponding 3 dB bandwidth of PZT 8 is totally out of range; with even higher $Q_m$, reliable measurement near $f_B$ is not possible for Standard $k_{33}$ mode of PIC 181.

As shown in experimental admittance values in Figure 7, considering peak values, the admittance value of PIC 181 (hard PZT) ranges from $10^{-6}$ S to near $3\times10^{-3}$ S, which corresponds to the range from $3.33\times10^2$ Ω to $10^6$ Ω, and the admittance value of PIC 255 (soft PZT) ranges from $10^{-5}$ S to $5\times10^{-4}$ S, which corresponds to the range from $2\times10^{-3}$ Ω to $10^5$ Ω. In accordance with Figure 8, The measurement error for PE samples made from hard PZT falls within 0.5 %, and that for samples made from soft PZT falls within 0.1 %. Compared to measurement error range out of scope for Standard Standard $k_{33}$ mode, the measurement error for PE samples less than 0.5 % is significantly smaller, and PE samples can effectively substitute Standard $k_{33}$ mode specimens for parameter determination purpose.

## 6. Conclusion

In this study, detailed parameter determination process using samples with PE configuration, along with simplified PE admittance equations and possible error causes, has been discussed. The center part of PE is likely to undergo fringing electric field and proper calibration is needed during fitting process. It was shown that the possible errors that may come from the volumes with canted polarization and statistical variation that comes from the parameters determined from Standard $k_{31}$ specimen are small (less than 0.2 %) enough to be neglected. With samples with PE configuration, researchers can obtain not only physical parameters with smaller statistical error by avoiding error propagation process, but also more reliable impedance/admittance curves from impedance analyzers. Accurate physical parameter values determined from PE are not only essential to elucidate heat dissipation mechanism of piezoelectric materials, but also necessary for accurate piezoelectric FEA simulation for prototype testing of piezoelectric devices. The physical meaning of the performance difference between hard and soft PZT will be report in the successive paper.

# Acknowledgement

This work was supported by Office of Naval Research under Grant Number N00014-17-1-2088

Supplementary Materials

for

**Physical Parameters and Loss Determination of Piezoceramic Using Partial Electrode: $k_{31}$ and $k_{33}$ mode cases**

Yoonsang Park[1,*], Hossein Daneshpajooh[1], Timo Scholehwar[2], Eberhard Hennig[2], and Kenji Uchino[1]

1) International Center for Actuators and Transducers (ICAT), The Pennsylvania State University, University Park, PA, 16802, USA

2) R&D Department, PI Ceramic GmbH, Lindenstrasse, 07589 Lederhose, Germany

## 1. ATILA FEA simulation conditions

### 1.1 FEA simulation of PE Geometry

The physical parameters of PZT 5A, which was already implemented in the software, were used for the simulation. 20 × 2.5 × 0.5 mm = length × width × thickness geometry for PE configuration was drawn with the center part that takes 10 % (2 mm) of the entire length. The mesh numbers of 20, 4, and 4 were assigned for length, width, and thickness, respectively, for side part; mesh numbers of 5, 4, and 4 were assigned for center part. Figure S1 shows mesh geometry for FEA simulation. Since the fundamental frequency corresponds to the vibration along the length direction (x direction in Figure S1), we assigned more meshes in length and less meshes in the other directions, to reduce simulation time while maintaining simulation accuracy. Table S1 shows physical parameters of PZT 5A used FEA simulation. It should be noted that ATILA considers loss isotropy, so losses are denoted without subscripted direction notation. 1 V was assigned to one of the upper one of two faces (orthogonal to thickness), and ground was assigned to another face.

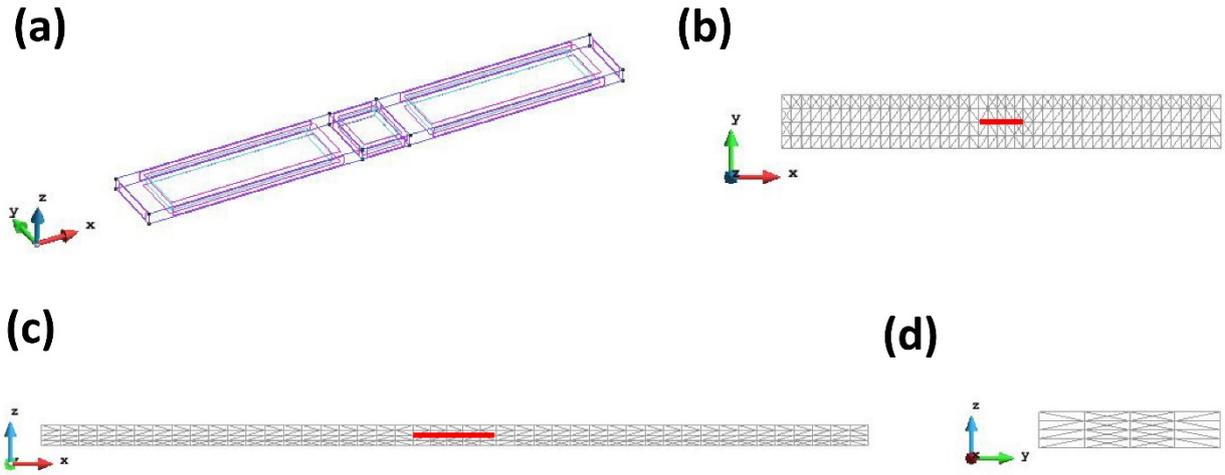

**Figure S1.** (a) PE geometry with coordinate in ATILA FEA and Mesh geometry for FEA simulation of PE configuration in (b) XY (c)XZ and (d) YZ plane. Red lines in (b) and (c) denote center part.

| Density (kg/m³) | 7750 | | |
|---|---|---|---|
| relative permittivity | $\varepsilon_{33}^X = 1700$ | dielectric loss | $\tan\delta = 0.02$ |
| elastic compliances (m²/N) | $s_{11}^E = 16.4 \times 10^{-12}$<br>$s_{33}^E = 18.8 \times 10^{-12}$ | elastic loss | $\tan\phi = 0.013$ |
| piezoelectric constant (pC/N) | $d_{31} = -171$<br>$d_{33} = 374$ | piezoelectric loss | $\tan\phi = 0.016$ |

**Table S1.** Physical parameter and loss values of PZT 5A used for FEA simulation

### 1.2 FEA Simulation of $k_{33}$ Standard Geometry

In order to obtain impedance values near antiresonance frequency and corresponding 3 dB bandwidth for various, commercially available piezoelectric ceramics, $k_{33}$ mode geometry with PZT 5A (soft), PZT 4 (semi-hard) and PZT 8 (hard) with the dimension of 20 × 5 × 5 mm = length × width × thickness (note that width and thickness have the same length). The mesh number was assigned to be 25,4 and 4 for length, width and thickness, respectively; the mesh geometry is schematically shown in Figure S2. 1 V was assigned to one of two faces orthogonal to the direction of polarization, and ground was assigned to the other face.

|  | PZT 4 | PZT 8 |
|---|---|---|
| mass density (kg/m³) | 7500 | 7600 |
| relative permittivity (…) | $\varepsilon_{33}^X = 1300$ | $\varepsilon_{33}^X = 1000$ |
| elastic compliance (m²/N) | $s_{33}^E = 15.5 \times 10^{-12}$ | $s_{33}^E = 13.9 \times 10^{-12}$ |
| Piezoelectric constant (pC/N) | $d_{33} = 289$ | $d_{33} = 218$ |
| dielectric loss | $\tan \delta = 0.004$ | $\tan \delta = 0.004$ |
| elastic loss | $\tan \phi = 0.002$ | $\tan \phi = 0.001$ |
| piezoelectric loss | $\tan \theta = 0.0035$ | $\tan \theta = 0.003$ |

**Table S2.** Physical parameters and loss values of PZT 4 and PZT 8 used for FEA simulation.

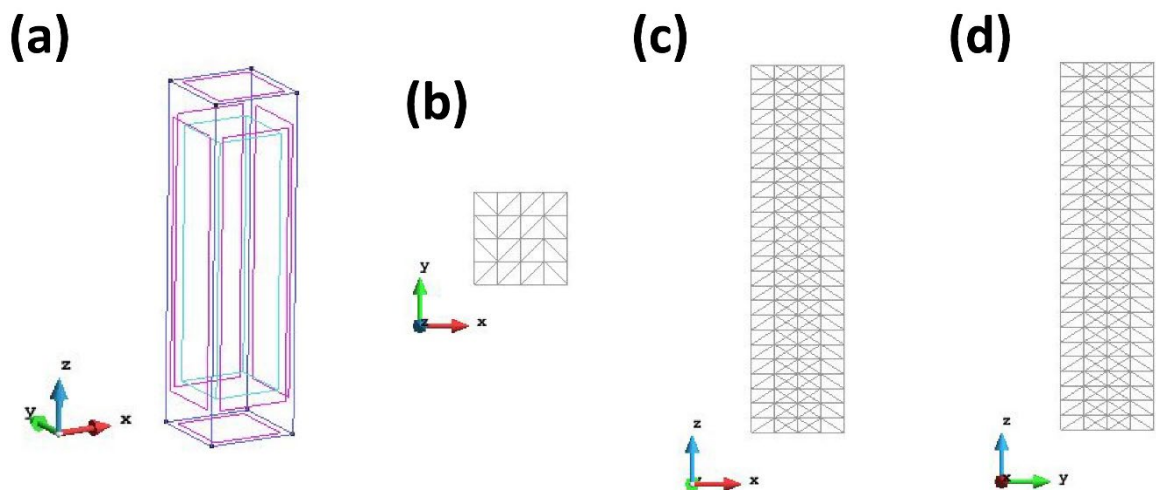

**Figure S2. (a)** PE geometry with coordinate in ATILA FEA and Mesh geometry for FEA simulation of PE configuration in **(b)** XY **(c)**XZ and **(d)** YZ plane.

## 2. Standard $k_{31}$ mode parameters

Standard $k_{31}$ mode samples made from PIC 181 and PIC 255 [PI Ceramic GmbH, Lederhose, Germany], with the dimension of 20 × 2.5 × 0.5 mm, were prepared. The off-resonance and near-resonance admittance/impedance spectra of each sample were measured with 4294A Precision Impedance Analyzer [Agilent Technologies, Santa Clara, CA]. Table S3 shows determined physical parameters and losses from 6 samples (for each material). The error denotes standard error of the mean taken from standard variation among each sample and error propagation method.

**Standard $k_{31}$ Parameters**

| PIC 181 | | | |
|---|---|---|---|
| Real Parameters | | | |
| $\varepsilon_{33}^X$ …. | $s_{11}^E$ (×10$^{-12}$ m²/N) | $d_{31}$ (pC/N) | $k_{31}$ …. |
| 1263 ± 3 | 11.94 ± 0.03 | -121.4 ± 0.4 | 0.332 ± 0.001 |
| Imaginary Parameters | | | |
| $\tan \delta_{33}'$ (%) | $\tan \phi_{11}'$ (%) | $\tan \theta_{31}'$ (%) | $\tan \chi_{31}$ (%) |
| 0.362 ± 0.003 | 0.054 ± 0.001 | 0.239 ± 0.004 | 0.10 ± 0.01 |
| PIC 255 | | | |
| Real Parameters | | | |
| $\varepsilon_{33}^X$ …. | $s_{11}^E$ (×10$^{-12}$ m²/N) | $d_{31}$ (pC/N) | $k_{31}$ …. |
| 1934 ± 3 | 16.77 ± 0.04 | -194.7 ± 0.5 | 0.363 ± 0.003 |
| Imaginary Parameters | | | |
| $\tan \delta_{33}'$ (%) | $\tan \phi_{11}'$ (%) | $\tan \theta_{31}'$ (%) | $\tan \chi_{31}$ (%) |
| 1.54 ± 0.01 | 1.21 ± 0.02 | 2.26 ± 0.04 | 1.8 ± 0.1 |

**Table S3.** Physical parameters and losses determined from Standard $k_{31}$ samples.

## 3. Experimental Admittance curves of PE samples

The full experimental admittance curves along with analytical fit of all types of PE specimens proposed in this study is shown in Figure S3. Note that the admittance curve agreement with analytical solution near resonance/antiresonance peak is in the main text.

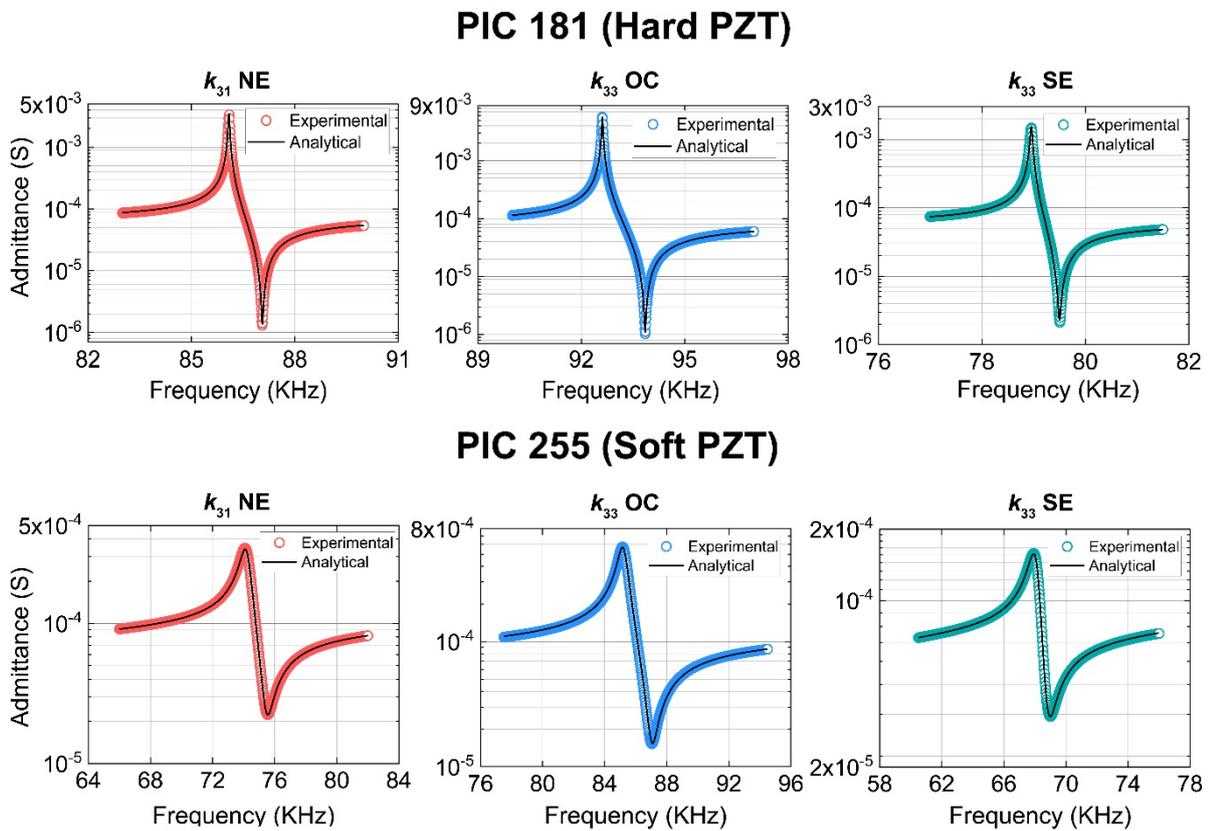

**Figure S3**. Full experimental admittance spectra of PE samples made of PIC 181 and PIC 255 measured with impedance analyzer. Black lines are analytical fitting curves.